\pdfoutput=1

\documentclass[11pt,twoside,a4paper,cmspaper,final,collab]{cms-tdr}
\def\svnVersion{40016:48625M}\def\cmsCernNo{2010-084}\def\cmsCernDate{\today}\def\cmsMessage{Submitted to Physics Letters B}

\begin{document}\cmsNoteHeader{SUS-10-003}

\hyphenation{had-ron-i-za-tion}
\hyphenation{cal-or-i-me-ter}
\hyphenation{de-vices}
\RCS$Revision: 29933 $
\RCS$HeadURL: svn+ssh://alverson@svn.cern.ch/reps/tdr2/papers/SUS-10-003/trunk/SUS-10-003.tex $
\RCS$Id: SUS-10-003.tex 29933 2011-01-08 22:48:05Z alverson $
\newcommand\rs{\raisebox{1.0ex}[-1.0ex]}
\newcommand{\ra}{\ensuremath{\rightarrow}}
\newcommand{\znunu}{\ensuremath{{\text Z} \ra \nu\bar{\nu}}}
\newcommand{\zmumu}{\ensuremath{{\text Z} \ra \mu\mu}}
\newcommand{\wmunu}{\ensuremath{{\text W} \ra \mu\nu}}
\newcommand{\wtaunu}{\ensuremath{{\text W} \ra \tau\nu}}
\newcommand{\dphi}{\ensuremath{\Delta \phi}}
\newcommand{\dphijj}{\ensuremath{\Delta \phi_{ j1,j2}}}
\newcommand{\Pt}{\ensuremath{{p_{\text T}}\xspace}}
\newcommand{\pts}{\ensuremath{p_{\text T}{\text s}}\xspace}
\newcommand{\Et}{\ensuremath{{E_{\text T}}\xspace}}
\newcommand{\ptjf}{\ensuremath{p_{\rm T}^{ {\rm j}_1} }}
\newcommand{\ptjs}{\ensuremath{p_{\rm T}^{ {\rm j}_2} }}
\newcommand{\ptjt}{\ensuremath{p_{\rm T}^{ {\rm j}_3} }}
\newcommand{\etajf}{\ensuremath{\eta^{ {\rm j}_1} }}
\newcommand{\etajs}{\ensuremath{\eta^{ {\rm j}_2} }}
\newcommand{\etajt}{\ensuremath{\eta^{ {\rm j}_3} }}
\newcommand{\ttj}{\ensuremath{\rm{t}\bar{\rm{t}} + jets}\xspace}
\newcommand{\wj}{\ensuremath{\rm W + jets}\xspace}
\newcommand{\zj}{\ensuremath{\rm Z + jets}\xspace}
\newcommand{\al}{\ensuremath{\alpha}}
\newcommand{\alt}{\ensuremath{\alpha_{\text{T}}}\xspace}
\newcommand{\etaabs}{\ensuremath{|\eta|}}
\newcommand{\mjj}{\ensuremath{M_{\text{inv}}^{j1,j2}}}
\newcommand{\chiznew}{\ensuremath{\chi^{0}}\xspace}
\newcommand{\chipnew}{\ensuremath{\chi^{+}}\xspace}
\newcommand{\sQuanew}{\ensuremath{\tilde{\rm q}}\xspace}
\newcommand{\sGlunew}{\ensuremath{\tilde{\rm g}}\xspace}
\newcommand{\ttNew}{\ensuremath{\rm{t}\bar{\rm{t}}}\xspace}
\newcommand{\tev}{\TeV}
\newcommand{\combIso}{Iso_{\textrm{comb.}}}
\renewcommand{\arraystretch}{1.2}
\newcommand{\bigNum}[2]{#1 \, \times \, 10 \, ^{#2}}

\newcommand{\raT}{\ensuremath{R_{\alt}}}
\newcommand{\RaT}{\ensuremath{R_{\alt}}\xspace}

\def\eslash{{\hbox{$E$\kern-0.6em\lower-.05ex\hbox{/}\kern0.10em}}}  
\def\vecmet{\mbox{$\vec{\eslash}_T$}} 
\def\vecet{\mbox{$\vec{E}_\text{T}$}} 
\def\MET{\mbox{$\eslash_\text{T}$}\xspace}
\def\met{\mbox{$\eslash_\text{T}$}\xspace}
\def\mex{\mbox{$\eslash_\text{x}$}} 
\def\mey{\mbox{$\eslash_\text{y}$}} 
\def\mepar{\mbox{$\eslash_\parallel$}}
\def\meperp{\mbox{$\eslash_\perp$}}
\def\Zmm{Z \rightarrow \mu\mu}
\def\metvec{\mbox{$\vec{\met}$}\xspace}
\def\metvecrec{\mbox{$\vec{\met}^{\rm rec}$}\xspace}
\def\metvecgen{\mbox{$\vec{\met}^{\rm gen}$}\xspace}
\def\metgen{\mbox{$\met^{\rm gen}$}\xspace}
\def\metparl{\mbox{$\mepar^{\rm rec}$}\xspace}
\def\metperp{\mbox{$\meperp^{\rm rec}$}\xspace}
\def\deltamet{\mbox{$\Delta\met$}\xspace}
\def\pthat{\mbox{$\hat{p}_T$}\xspace}
\def\hslash{{\hbox{$H$\kern-0.8em\lower-.05ex\hbox{/}\kern0.10em}}}
\def\MHT{\mbox{$\hslash_\text{T}$}\xspace}
\def\mht{\mbox{$\hslash_\text{T}$}\xspace}
\def\sumet{\mbox{$\sum \rm{E}_\text{T}$}\xspace}
\def\scalht{\mbox{$H_\text{T}$}\xspace}
\def\etmiss{\mbox{$\eslash_\text{T}$}\xspace}
\def\htmiss{\mbox{$\hslash_\text{T}$}\xspace}
\def\mtt{\mbox{$\rm{M}_\text{T2}$}\xspace}
\def\rmec{\mbox{$R_{\mht/\met}$}\xspace}
\def\bdphi{\mbox{$\Delta\phi^{*}$}\xspace}
\def\bigeslash{{\hbox{$E$\kern-0.38em\lower-.05ex\hbox{/}\kern0.10em}}}  
\def\bigmet{\mbox{$\bigeslash_T$}}
\def\bighslash{{\hbox{$H$\kern-0.6em\lower-.05ex\hbox{/}\kern0.10em}}}  
\def\bigmht{\mbox{$\bighslash_T$}} 
\def\incl{\includegraphics[width=0.49\linewidth]}
\def\inclrot{\includegraphics[angle=90,width=0.47\linewidth]}
\def\INCL{\includegraphics[angle=90,width=0.45\linewidth]}
\def\Incl{\includegraphics[angle=90,width=0.60\linewidth]}

\cmsNoteHeader{SUS-10-003} 
\title{Search for Supersymmetry in pp Collisions at 7 TeV in Events with Jets and Missing Transverse Energy}

\author[cern]{The CMS Collaboration}

\date{\today}

\abstract{

A search for supersymmetry with R-parity conservation in
proton-proton collisions at a centre-of-mass energy of 7 TeV is
presented. The data correspond to an integrated luminosity of
35~pb$^{-1}$ collected by the CMS experiment at the LHC.
The search is
performed in events with jets and significant missing transverse
energy, characteristic of the decays of heavy, pair-produced squarks
and gluinos.  The primary background, from standard model multijet
production, is reduced by several orders of magnitude to a negligible
level by the application of a set of robust kinematic requirements.
With this selection, the data are consistent with the standard model
backgrounds, namely t$\bar{\rm t}$, W + jet and Z + jet production,
which are estimated from data control samples.
Limits are set on the parameters
of the constrained minimal supersymmetric extension of the standard
model. These limits extend those set previously by experiments
at the Tevatron and LEP colliders.\\

}

\hypersetup{%
pdfauthor={CMS Collaboration},%
pdftitle={Search for Supersymmetry in pp Collisions at 7 TeV in Events with Jets and Missing Transverse Energy},%
pdfsubject={CMS},%
pdfkeywords={CMS, physics, supersymmetry}}

\maketitle 

\section{Introduction}

The standard model (SM) of particle physics has been extremely 
successful in describing phenomena at the highest energies attained thus far. 
Nevertheless, it is widely believed to be only an effective
description of a more complete theory, which supersedes it at higher
energy scales. Of particular theoretical interest is supersymmetry
(SUSY)~\cite{ref:SUSY-1,ref:SUSY0,ref:SUSY1,ref:SUSY2,ref:SUSY3, ref:SUSY4}, which
solves the ``hierarchy problem''~\cite{ref:hierarchy1,ref:hierarchy2}
of the SM at the expense of introducing a large number of
supersymmetric particles with the same quantum numbers as the SM
particles, but differing by half a unit of spin. If R-parity
conservation \cite{Farrar:1978xj} is assumed, supersymmetric particles are produced in
pairs and decay to the lightest supersymmetric particle (LSP).
If the LSP is neutral and weakly-interacting, it goes undetected 
giving rise to a signature with missing energy.

Experiments at the energy frontier, i.e. at the Fermilab Tevatron 
collider~\cite{CDFLimits,CDFtrileptons,D0Limits,Abazov200934} and previously 
at the CERN $\rm{Sp\bar{p}S}$~\cite{UA1Limits,UA2Limits}, 
HERA~\cite{HERALimits1,HERALimits2} and LEP~\cite{LEPLimits} colliders, 
have performed extensive searches for signs of SUSY. 
In the absence of a positive signal, lower limits on 
the masses of SUSY particles have been set.
With its higher centre-of-mass energy of 7\tev, the Large Hadron Collider (LHC) at CERN
could produce SUSY particles (sparticles) with masses larger
than the current limits.  The dominant
production channels of heavy coloured sparticles at the LHC are
squark-squark, squark-gluino and gluino-gluino pair production.  In
the context of SUSY with R-parity conservation, heavy squarks and
gluinos decay into quarks, gluons and other SM particles, as well as a
neutralino (i.e.~the LSP), which escapes undetected, leading to final states with
several hadronic jets and large missing transverse energy.  While squark-squark
production usually leads to two jets, gluino production typically
results in higher jet multiplicities.  This Letter describes a 
search for the production and decay of SUSY particles by the CMS
experiment, in events with two or more energetic jets and significant
imbalance of transverse energy.  

The search is not optimized in the context of any particular 
model of SUSY. To interpret the results, a simplified 
and practical model of SUSY-breaking, the constrained minimal 
supersymmetric extension of the standard model (CMSSM)~\cite{ref:CMSSM, ref:MSUGRA}, 
is used.  The CMSSM is described by five parameters: the universal scalar 
and gaugino mass parameters ($m_0$ and $m_{1/2}$, respectively), 
the universal trilinear soft SUSY breaking parameter $A_0$, and two low-energy parameters, 
the ratio of the two vacuum expectation values of the two Higgs doublets, 
$\tan\beta$, and the sign of the Higgs mixing parameter, $\sign(\mu)$. 
Throughout the Letter, two CMSSM parameter sets, referred to as LM0 and 
LM1~\cite{PAS-SUS-09-001}, are used to illustrate possible CMSSM yields. 
The parameter values defining LM0 are $m_0=200\gev$, $m_{1/2}=160\gev$, 
$A_0=-400\gev$, $\tan\beta=10$, and $\sign(\mu)>0$. Those for LM1 
are $m_0=60\gev$, $m_{1/2}=250\gev$, $A_0=0$, $\tan\beta=10$, 
and $\sign(\mu)>0$.

\section{The CMS Detector}

The central feature of the CMS apparatus is a superconducting solenoid, 13~m in length and
6~m in diameter, which provides an axial magnetic field of 3.8~T. The bore of the solenoid
is instrumented with various particle detection systems.  The steel return yoke outside the
solenoid is in turn instrumented with gas detectors used to identify muons.
Charged particle trajectories are measured by the silicon pixel and strip tracker,
with full azimuthal coverage within $|\eta| <$ 2.5, where the pseudorapidity $\eta$ is defined
as $\eta = -\ln \tan (\theta/2)$, with $\theta$ being the polar angle of the trajectory of
the particle with respect to the counterclockwise beam direction.
A lead-tungstate crystal electromagnetic calorimeter (ECAL) and a brass/scintillator hadron
calorimeter (HCAL) surround the tracking volume and cover the region $|\eta| < 3$.
In the region $\vert \eta \vert<  1.74$, the HCAL cells have widths of
0.087 in pseudorapidity and 0.087 in azimuth ($\phi$). In the
$(\eta,\phi)$ plane, and for $|\eta|< 1.48$, the HCAL cells map on to $5
\times 5$ ECAL crystal arrays to form calorimeter towers projecting
radially outwards from close to the nominal interaction point. At larger
values of $\vert \eta \vert$, the size of the towers increases and the
matching ECAL arrays contain fewer crystals. Within each tower, the
energy deposits in ECAL and HCAL cells are summed to define the
calorimeter tower energies, subsequently used to provide the energies
and directions of hadronic jets.
The detector is nearly hermetic, which allows for energy-balance measurements in the plane
transverse to the beam axis. A more detailed description of the CMS detector can be found
elsewhere \cite{ref:CMS}.

\section{Event Selection \label{sec:evtsel}}

\subsection{Hadronic final state selection \label{sec:hadsel}}

The data sample used in this analysis is recorded with a trigger based on the scalar sum of the transverse energy \Et~of jets, defined in general as $\scalht= \sum_{i=1}^{{\rm N_{jet}}} \Et^{ {\rm j}_i}$, where ${\rm N_{jet}}$ is the number of jets.  Events are selected if they satisfy $\scalht^{\rm trigger} > 150\gev$.

At the trigger level, the calorimeter response is not corrected to achieve a uniform and absolute scale of transverse jet energy; nevertheless the trigger requirement is fully efficient for events with an offline-reconstructed \scalht in excess of $250\gev$, thus providing a high signal efficiency for the region of the CMSSM parameter space relevant for the present search, where squarks and gluinos have masses of several hundred GeV.
Additionally, events are required to have at least one
good reconstructed pp interaction vertex~\cite{PAS-TRK-10-005}.

Jets are reconstructed offline from the energy deposits in the calorimeter towers, clustered by the  anti-k$_\text{T}$ algorithm~\cite{ref:antikt} with a size parameter of $0.5$. In this process, the contribution from each calorimeter tower is assigned a momentum, the magnitude and direction of which are given by the energy measured in the tower and the coordinates of the tower.  The raw jet energy is obtained from the sum of the tower energies, and the raw jet momentum by the vectorial sum of the tower momenta, resulting in a nonzero jet mass. The raw jet energies are corrected to establish a relative uniform response of the calorimeter in $\eta$ and a calibrated absolute response in transverse momentum \pt.
The uncertainty on the energy scale of these
corrected jets varies between $3\%$ and $5\%$, depending on the jet
\pt and $|\eta|$~\cite{PAS-JME-10-010}. The jets considered in this
analysis are required to have $\Et>50\gev$,
$|\eta| < 3$ and to pass jet identification criteria~\cite{JME-09-008}
designed to reject spurious signals in the calorimeters. The
pseudorapidity of the jet with the highest \Et~(leading jet) is required to be within
$|\eta|<2.5$ and the transverse energy of each of the two leading
jets must exceed 100\gev.

Events with jets passing the \Et\ threshold but not satisfying the jet
identification criteria or the $\eta$ acceptance requirement are vetoed, as
this deposited energy is not accounted for in the event kinematics. Similarly,
events in which an isolated lepton (electron~\cite{PAS-EGM-10-004}
or muon~\cite{PAS-MUO-10-002}) with $\pt > 10\gev$ is identified are
rejected to suppress events with genuine missing energy from neutrinos. Furthermore,
to select a pure multi-jet topology, events are vetoed in which an isolated
photon~\cite{PAS-EGM-10-005} with $\pt > 25\gev$ is found. These vetoes
reject 5\% of the previously selected events in data and simulation.

\begin{figure}[t]
   \begin{center}
   \incl{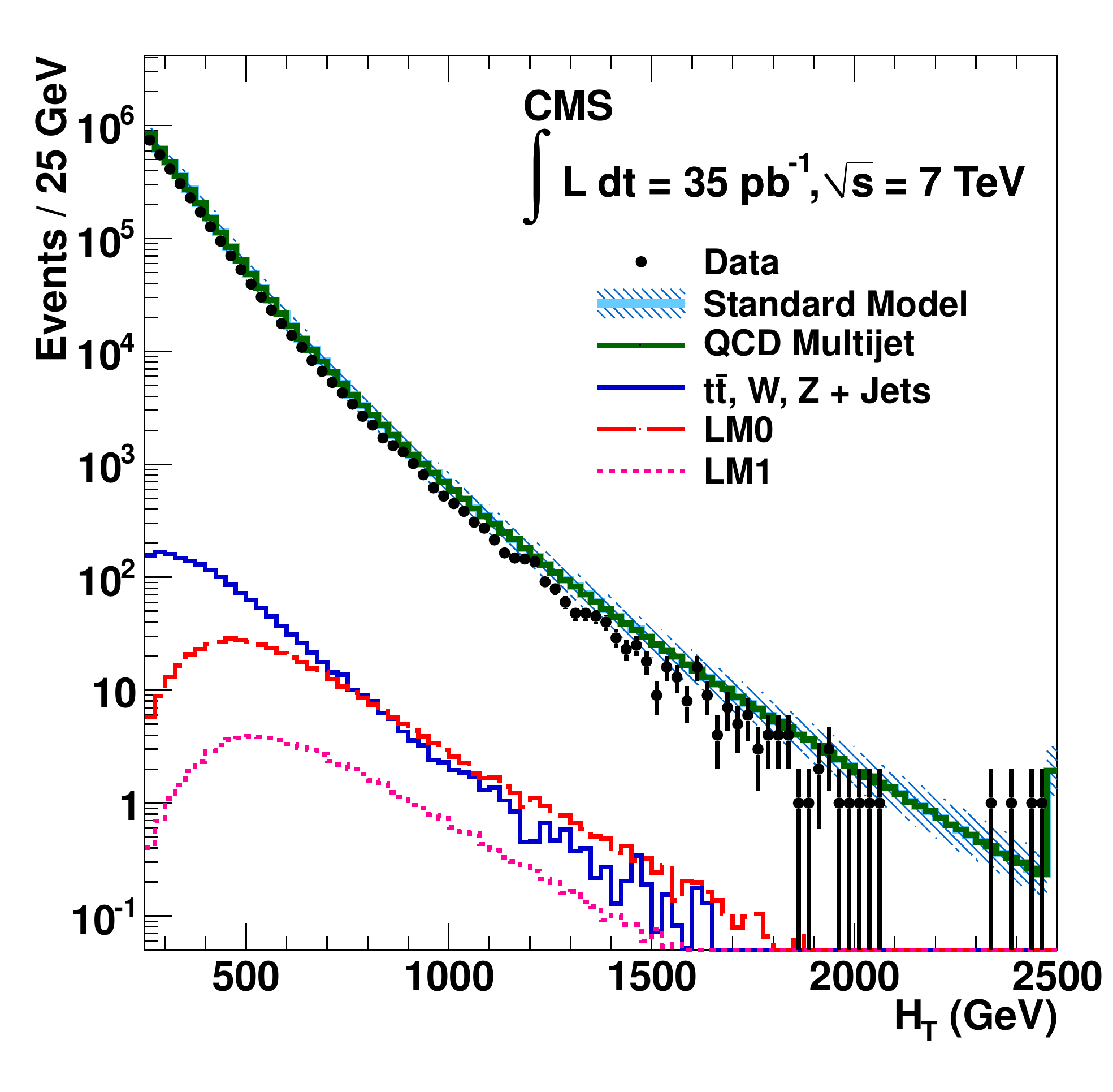}
   \caption{\label{fig:ht} $\scalht$ distribution after preselection,
     for data as well as for all standard model backgrounds and two SUSY
     signal samples with parameter sets LM0 and LM1, normalized to an
     integrated luminosity of 35~pb$^{-1}$. The hatched area corresponds to the uncertainty
     in the SM estimate as defined in Section~\ref{sec:hadsel}. The SM distributions are
     only displayed for illustration purposes, as they are the result of Monte
     Carlo simulation, while the actual estimate of the background from SM
     processes in this search is based on data, as described in detail
     in Section~\ref{sec:background}.
}
\end{center}
\end{figure}

At this preselection stage, the background from multijet production, as predicted by quantum chromodynamics (QCD), is still several orders of magnitude larger than the typical signal expected from SUSY.  The \scalht distribution for the
selected events is shown in Fig.~\ref{fig:ht} and compared to simulation-based background estimates.
The QCD multijet background is estimated using the \PYTHIA6.4~\cite{pythia} Monte Carlo generator with
tune Z2~\cite{Z2}. Electroweak backgrounds from W + jets, \znunu\ + jets and \ttNew + jets
events, which will be referred to collectively as the electroweak (EWK) backgrounds
in what follows, are simulated using \MADGRAPH~\cite{madgraph}.
The SM distribution, i.e.\ the sum of the QCD multijet and EWK distributions, is indicated in Fig.~\ref{fig:ht}
as a hatched band representing the combined statistical and systematic uncertainties from the jet energy scale
and resolution. The expected \scalht distributions for two low-mass SUSY
signal points, LM0 and LM1, are overlaid. With the exception of \ttNew,
the SM processes fall off exponentially over the entire \scalht range,
whereas a broad peak at values of a few hundred of GeV is expected for
the signal models.  The selection is tightened by requiring the \scalht
of all jets to exceed 350\gev, thus ensuring large hadronic activity
in the event. This requirement substantially reduces the contributions
from SM processes while maintaining a high efficiency for the SUSY
topologies considered.

\subsection{Final event selection for SUSY search \label{sec:finalsel}}

Jet mismeasurements, caused by possible detection inefficiencies or by nonuniformities in the
calibration of the calorimeters, are the dominant source of large missing transverse energy \met
in events from QCD multijet production.
To control this background and to separate it from a
genuine missing energy signal, a variable that is robust against
energy mismeasurements, \alt, is used.  For events with two jets,
\alt, first introduced in Refs.~\cite{PAS-SUS-08-005,PAS-SUS-09-001} and
inspired by Ref.~\cite{Randall:2008rw}, is defined as

\begin{equation*}
\label{eq:altsimple}
\alt = \Et^{\rm j_2}/M_\text{T},
\end{equation*}

where $\Et^{\rm j_2}$ is the transverse energy of the less energetic of the two jets
in the event and $M_\text{T}$ is the transverse mass of the di-jet system, defined as

\begin{equation*}
M_\text{T} = \sqrt{ \left( \sum_{i=1}^2
\Et^{{\rm j}_i} \right)^2 - \left( \sum_{i=1}^2 p_x^{{\rm j}_i} \right)^2 - \left(
\sum_{i=1}^2 p_y^{{\rm j}_i} \right)^2}.
\end{equation*}

For a perfectly measured di-jet event, with $\Et^{\rm j_1} = \Et^{\rm j_2}$ and
jets back to back in $\phi$, and in the limit where the
jet momenta are large compared to their masses, the value of \alt is
0.5. In the case of an imbalance in the measured transverse
energies of back to back jets, \alt takes on values smaller than
0.5, while for jets that are not back to back, \alt can be greater
than 0.5.

For larger jet multiplicities, the $n$-jet system is reduced to a
di-jet system by combining the jets in the event into two
pseudo-jets. The $\Et$ of each of the two pseudo-jets is calculated
as the scalar sum of the contributing jet $\Et$'s.  The combination
chosen is the one that minimizes the $\Et$ difference between the two
pseudo-jets. This simple clustering criterion has been found to
result in the best separation between QCD multijet events and events with
genuine \met.

Values of $\alpha_{\rm T}$ above 0.5 can occur for QCD multijet events,
either with multiple jets failing the $\Et>50\gev$  requirement,
or with missing transverse energy arising from jet energy resolution or
severe jet energy under-measurements due to detector inefficiencies.
On the other hand, events with genuine \met often have much larger values of \alt, resulting
in a good separation of signal events from the QCD multijet background.

\begin{figure}[t]
   \begin{center}
   \incl{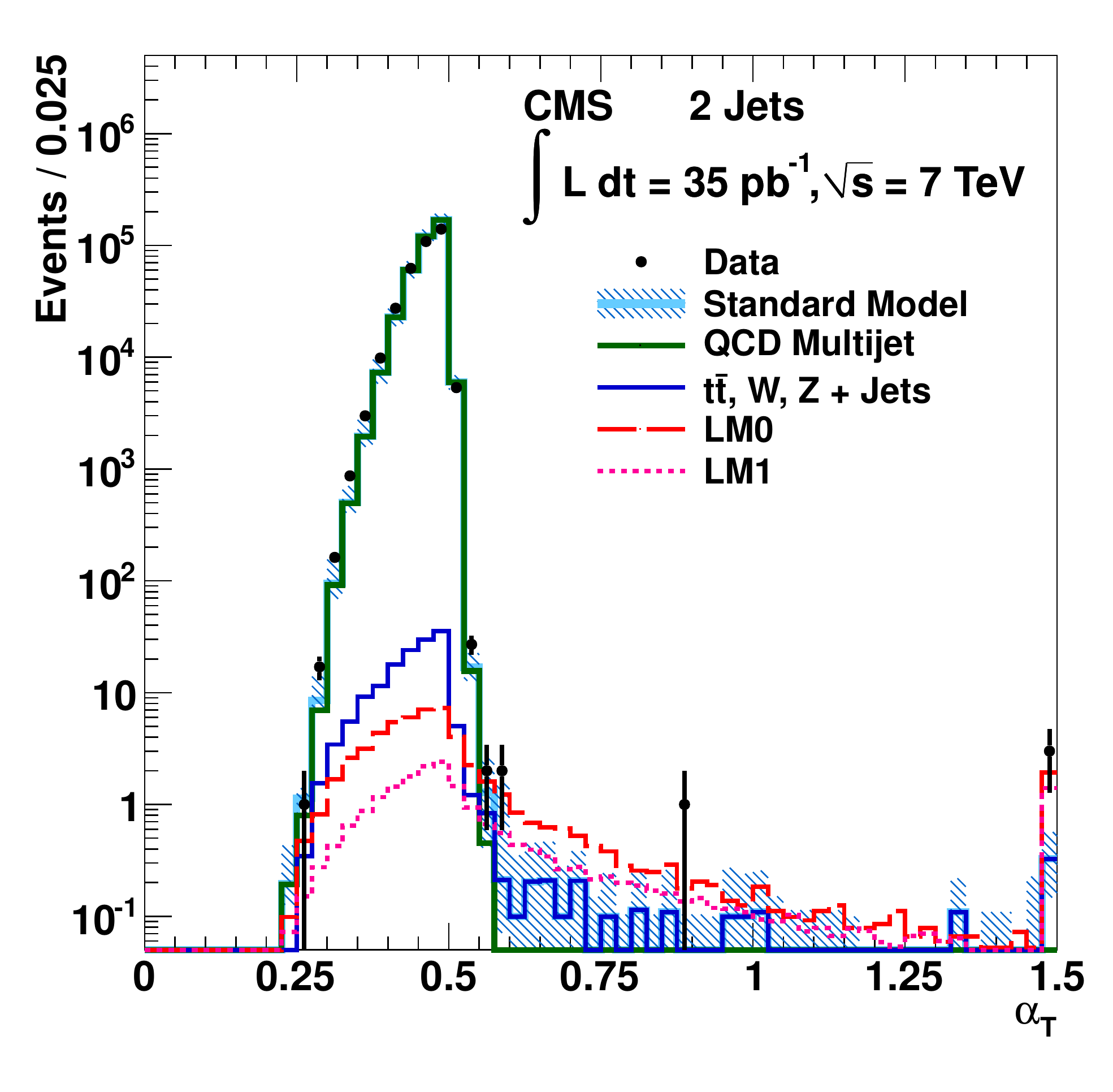}
   \incl{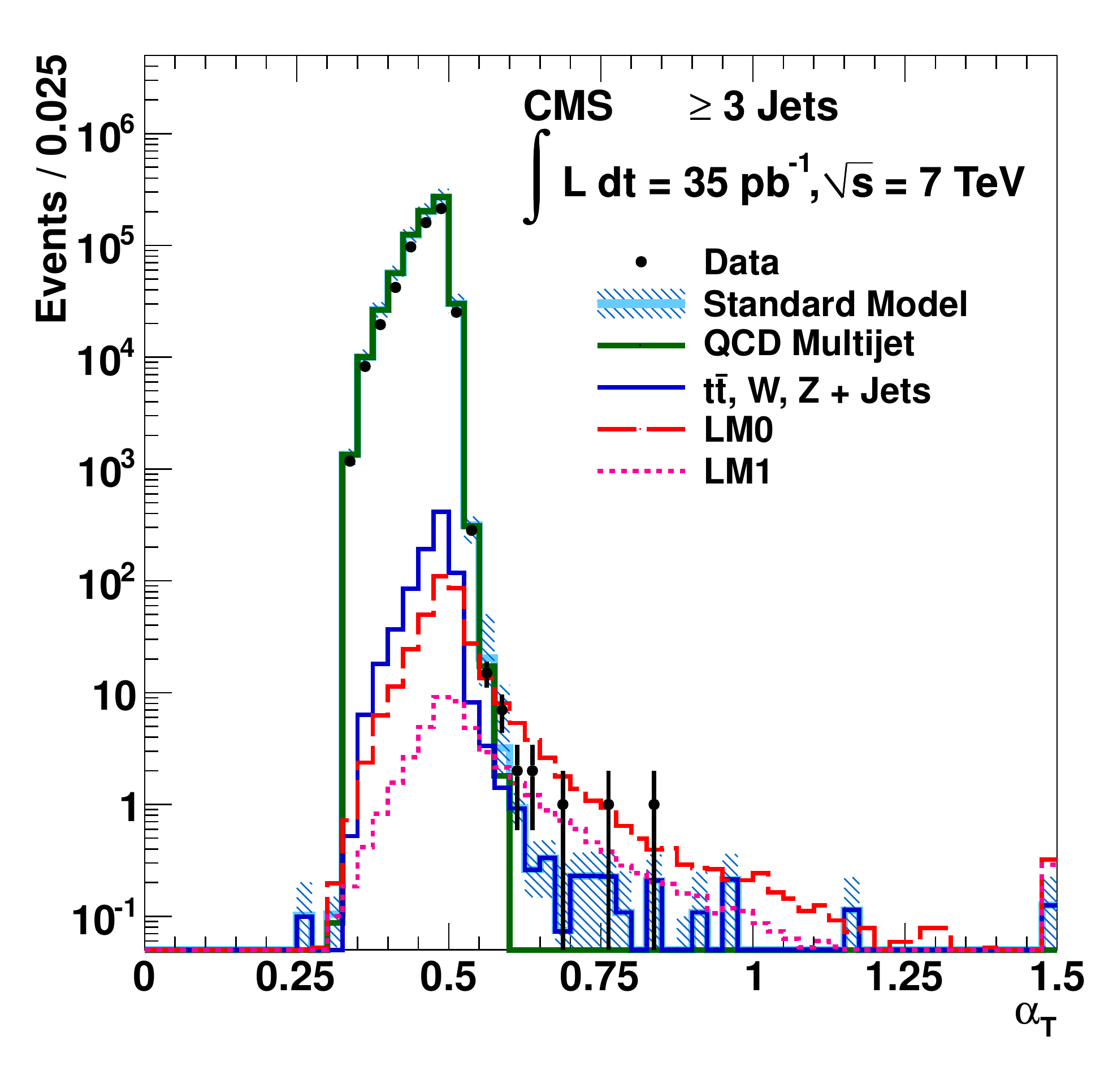}
 \caption{\label{fig:altbare} Distribution of \alt for di-jet events (left) and $\geq 3$-jet events (right), requiring $\scalht > 350\gev$. Events with $\alt > 1.5$ are included in the rightmost bin. In both figures the hatched area corresponds to the uncertainty in the SM estimate as defined in Section~\ref{sec:hadsel}.}
\end{center}
\end{figure}

The \alt distributions are shown separately for di-jet and $\geq$ 3-jet
events in Fig.~\ref{fig:altbare}.  As anticipated, these distributions peak at $\alt=0.5$ for
QCD multijet events and then fall sharply in the range $0.5$ to $0.55$, reaching a level 4--5 orders of
magnitude lower than the peak value.  Multijet events
from QCD background are therefore efficiently rejected by requiring
\alt to exceed 0.55. Given the selection requirement $\scalht>350\gev$, this threshold on $\alt$
is equivalent to demanding $\mht/\scalht>0.4$, i.e.to $\mht>140\gev$.

To reject events with false missing energy arising from significant jet 
mismeasurements in masked regions of the ECAL, which amount to about 1\%
of the ECAL channel count, the following procedure is employed. 
The jet-based estimate of the missing transverse energy,
$\mht = |\vec{\mht}|=| - \sum_{\rm  jets} \vec{\pt}_{\rm jet}|$, which is obtained
by summing the transverse momenta of all the jets in the event,
is now recomputed while ignoring one of the reconstructed jets.  The difference in 
azimuth between the recomputed $\vec{\mht}$ and the ignored jet is then calculated.
The $\vec{\mht}$ is recomputed for each configuration that results
from ignoring, in turn, each of the jets in the event, while leaving all other 
jets intact, and the minimum of all the azimuthal differences, $\Delta\phi^*$, is found.  
The jet whose subtraction from the calculation $\vec{\mht}$ yields this minimum value,
is identified as the jet that is most likely to have given rise to the $\mht$ in the event.
Events with $\Delta\phi^* < 0.5$ are rejected if the
distance in the ($\eta,\phi$) plane between the selected jet and the
closest masked ECAL region, $\Delta R_{\rm ECAL}$, is smaller than 0.3.

Artificially large values of \mht can also result in events with multiple 
jets below the selection requirement of $\Et>50\gev$, since these jets are
not included in the computation of \mht.  To protect against these events,   
\mht, i.e. the jet-based estimate of the missing energy, \mht, is compared to
the calorimeter tower-based estimate, $\met^{\rm calo}$, which includes
the energy from all jets, irrespective of threshold~\cite{PAS-JME-10-005}.
Events with $R_{\rm miss}=\mht/\met^{\rm calo} > 1.25$ are rejected.

Table~\ref{tab:smallyields} lists the number of events passing
each step of the event selection for data and simulation.  The
expectations from simulation are listed only for comparison;
the actual expected yields from standard model processes are
determined from control data samples, as described in the
following section.

After the selection requirements on \alt, $\Delta R_{\rm ECAL}$ and $R_{\rm miss}$, the QCD
multijet background predicted by \PYTHIA6.4 is less than one event for an integrated luminosity of 35~pb$^{-1}$. This estimate
is also obtained with  \PYTHIA8.1~\cite{pythia8} (tune 1) and with the
\MADGRAPH generator. After all selection requirements, the only significant remaining background
stems from electroweak processes with genuine \met in the final state.  In the di-jet case, the largest
backgrounds with real missing energy are the associated production of
W or Z bosons with jets, followed by the weak decays \znunu\ and
\wtaunu, or by leptonic W/Z decays in which one or more leptons are not
reconstructed. At higher jet multiplicities, \ttNew production followed by
semileptonic weak decays of the t and $\bar{\rm t}$ quarks becomes
important. In this case, the three backgrounds, \znunu\ + jets, W + jets and
\ttNew, are of roughly equal size.
The largest fraction of the W + jets and \ttNew backgrounds stem
from \wtaunu\ decays where in two thirds of the cases the $\tau$
decays hadronically and is identified as a jet. The two remaining backgrounds from electrons or muons produced in W decays
that fail either the isolation or acceptance requirements ($\pt>10\gev$ and
$\eta$ coverage) are of similar size.

\begin{table*}[htb]
\caption{The number of events observed and expected from Monte Carlo simulation after the selection requirements, for data and background samples (QCD multijet simulated with \PYTHIA6.4(Z2), \znunu, W~+jets, \ttNew). The quoted errors represent the statistical uncertainties on the yields and all numbers are normalized to an  integrated luminosity of 35~pb$^{-1}$.}
\label{tab:smallyields}
\begin{center}
\begin{tabular}{|l|c|c|c|c|c|c|}
\hline
Selection              & Data  & SM           & QCD multijet & \znunu & \wj          & \ttNew \\
\hline
$\scalht > 250\gev$      & 4.68M & 5.81M        & 5.81M         & 290 & 2.0k & 2.5k  \\
$ \Et^{\rm j_2}> 100\gev$   & 2.89M & 3.40M        & 3.40M         & 160  & 610   & 830 \\
$\scalht > 350\gev$      & 908k  & 1.11M        & 1.11M         & 80  & 280 & 650 \\
\hline
$\alpha_T > 0.55$            & 37    & 30.5$\pm$4.7 & 19.5$\pm$4.6  & 4.2$\pm$0.6   & 3.9$\pm$0.7    & 2.8$\pm$0.1 \\
$\Delta R_{\rm ECAL} > 0.3 \vee \dphi^* > 0.5$ & 32    & 24.5$\pm$4.2 & 14.3$\pm$4.1  & 4.2$\pm$0.6   & 3.6$\pm$0.6    & 2.4$\pm$0.1 \\
$R_{\rm miss}$ $< 1.25$      & 13    & 9.3$\pm$0.9 & 0.03$\pm$0.02 & 4.1$\pm$0.6   & 3.3$\pm$0.6    & 1.8$\pm$0.1 \\
\hline
\end{tabular}
\end{center}
\end{table*}

\section{Background Estimate from Data\label{sec:background}}

The SM background in the signal region is estimated directly from data
using two independent methods. The first method makes use of control
regions at lower \scalht to estimate the total background from all SM
processes (Section~\ref{sec:inclusive-bkg}), while the second method
estimates the contribution from electroweak processes using \wmunu\ +
jets (Section~\ref{sec:w-bkg}) and $\gamma$ + jets
(Section~\ref{sec:z-bkg}) events in the data.

\subsection{Inclusive background estimate \label{sec:inclusive-bkg}}

The total background can be estimated from two control regions at low
\scalht: the HT250 region, which contains events with \scalht between
250 and 300\gev, and the HT300 region, which contains events with
\scalht between 300 and 350\gev. Given the current experimental limits on
the squark and gluino masses, these two regions are expected to be
dominated by SM processes.  The search region for the signal, which is
referred to as the HT350 region in what follows, is defined as events
with $\scalht>350\gev$.

The method is based on the variable \RaT, defined as the ratio of the number of
events passing and failing a requirement on \alt, given all other selection
requirements.  To minimize the efficiency bias arising from the phase space reduction
in the lower \scalht regions, the \pt thresholds for the two bins are adjusted to keep the ratio of \pt/\scalht constant in each region.
In the HT300 region, the resulting thresholds are 86 GeV for the two leading
jets and 43 GeV for additional jets.  In the HT250 region the respective thresholds
are 71 and 36\gev.  In the absence of a SUSY signal, the ratio \RaT can then be
extrapolated from the measured values in both control regions to
predict the value in the signal region, HT350.

\begin{figure}[!h]
  \begin{center}
      \includegraphics[angle=0,width=0.45\textwidth]{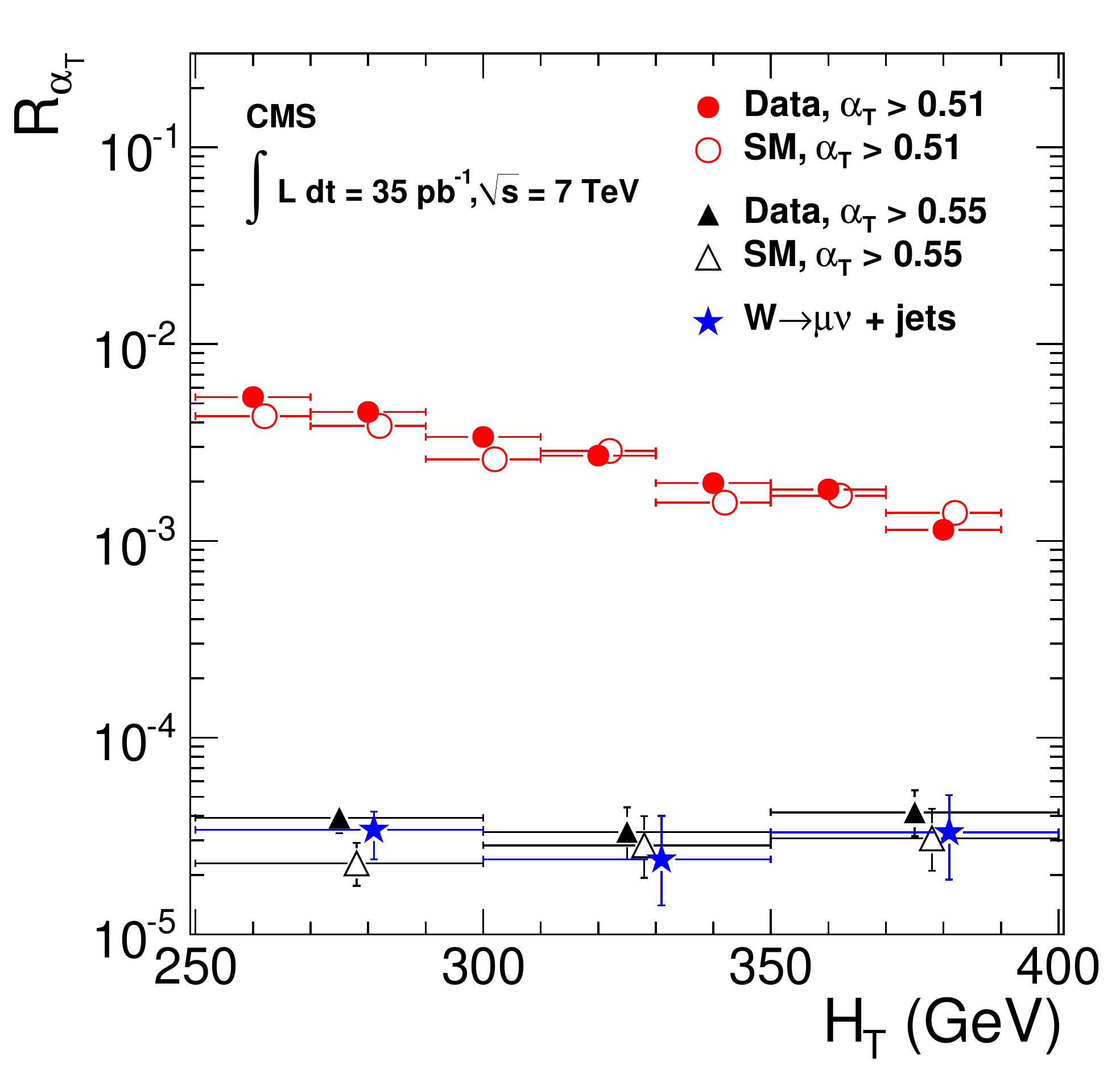}
     \includegraphics[angle=0,width=0.45\textwidth]{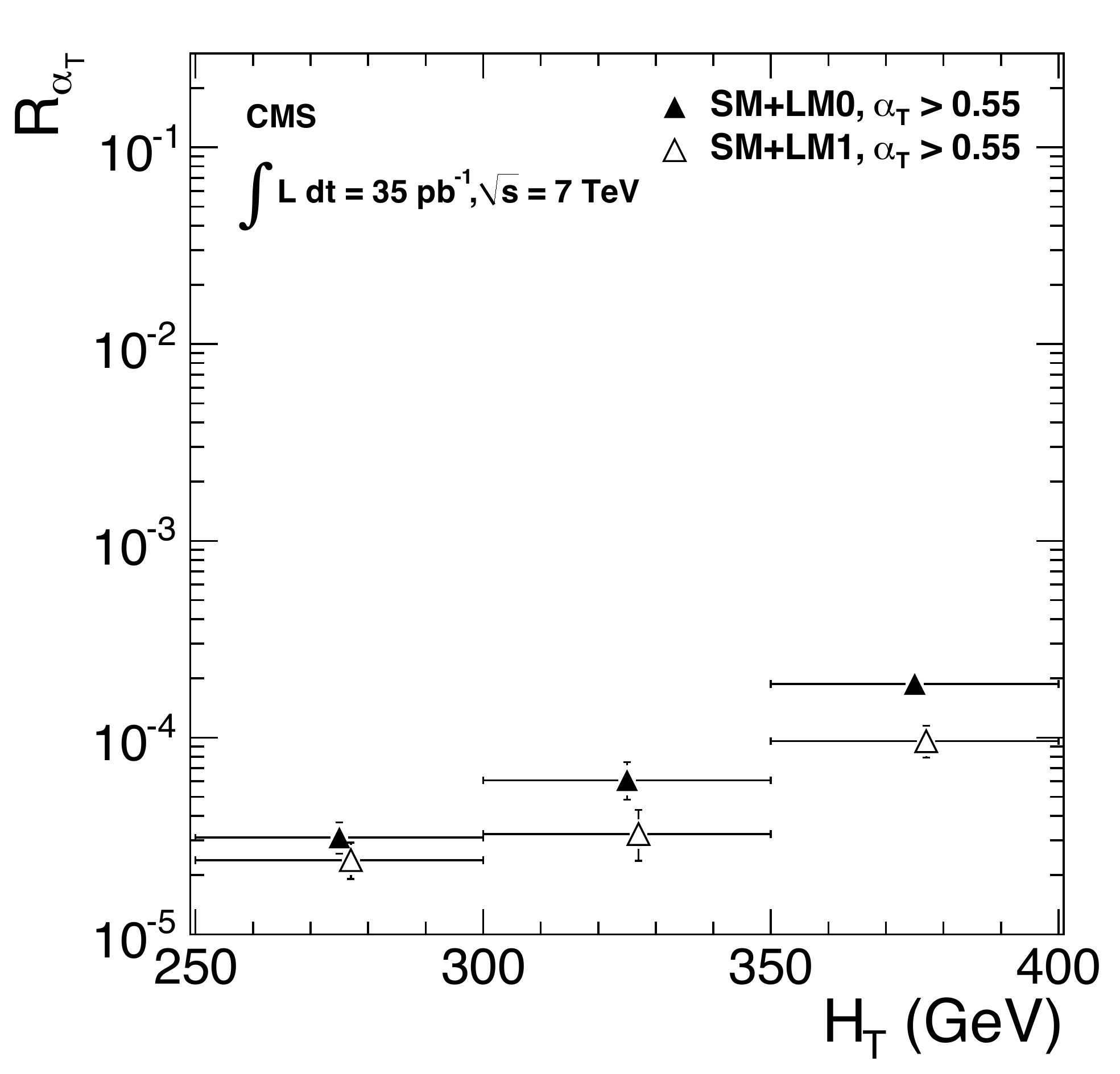}
  \end{center}
  \caption{\label{fig:ratio_vs_ht} Evolution of the ratio \RaT as a function of \scalht for events with $N_{\mathrm{jet}}\geq2$; (left) for data and SM backgrounds, and two different values of \alt, as well as for an independent \wmunu\ + jets control sample (Section~\ref{sec:w-bkg}); (right) for the SM backgrounds added to the SUSY signal expected from each of the two benchmark points, LM0 and LM1.
  Markers are offset horizontally for improved visibility.
}
\end{figure}

Figure~\ref{fig:ratio_vs_ht} (left) shows the evolution of \RaT as a function
of \scalht for two thresholds on $\alt$, namely $\alt > 0.51$ and $\alt > 0.55$.  For
$\alt > 0.51$, the numerator of \RaT is dominated by the QCD multijet
background, for which the missing transverse energy mostly originates
from energy mismeasurements. As the relative resolution of
calorimetric energy measurements improves with energy, and therefore
with \scalht, the relative importance of this background is expected
to decrease with increasing \scalht. This effect is clearly visible in
Fig.~\ref{fig:ratio_vs_ht} (left), which shows the falling behaviour for seven equidistant bins in \scalht.  In contrast, for $\alt > 0.55$, the
numerator of \RaT is dominated by the electroweak background, with
genuine \met, the relative importance of which is expected to be
constant with increasing \scalht.

The latter behaviour is confirmed in an independent sample of events
with a W decaying to $\mu\nu$, accompanied by jets. (The selection of
these events is given in Section~\ref{sec:w-bkg}.) The ratio of the
number of selected \wmunu\ + jets events to the number of events
failing the W selection (hence dominated by the QCD multijet background) is shown
in Fig.~\ref{fig:ratio_vs_ht} (left) for the same \scalht bins, and confirms
the independence of \RaT on \scalht when the numerator is dominated by
events with genuine \met.

The ratio \RaT in the HT350 region can be estimated from the \RaT values
measured in regions HT250 and HT300, for $\alt > 0.55$, using the double ratio $R_{\rm R}$:

\begin{equation}
\label{eqn:RR}
  R_{\mathrm{R}} =
  \frac{\RaT(\rm HT300)}{\RaT(\rm HT250)} =
  \frac{\RaT(\rm HT350)}{\RaT(\rm HT300)}.
\end{equation}

The total number of events with $\alt > 0.55$ expected from SM
processes in the signal region is the product of the
extrapolated $\raT$(HT350) and the number of events with $\alt<0.55$
in the $\scalht>350\gev$ region. The total number of background events in HT350
thus estimated is $9.4 ^{+4.8}_{-4.0}$ (stat) $\pm$1.0 (syst). The
dominant systematic uncertainty for this method is estimated by
varying the relative magnitude of the three EWK processes,
while maintaining the \scalht dependence of each process to the one 
predicted by the Monte Carlo.
The background estimate is insensitive to this variation:
since $R_{\mathrm{R}} \approx 1$, the change in the estimate is always
much smaller than the statistical uncertainty. Even under extreme variations
of the individual EWK processes by up to five times higher values than those
predicted by simulation, the systematic uncertainty is at least a factor two
smaller than the statistical uncertainty. The same comments hold for the background estimate
variants described below.

For comparison, Fig.~\ref{fig:ratio_vs_ht} (right) shows the
expectation of \RaT if a SUSY signal from each of the
two benchmark points LM0 and LM1 were present in addition to the
SM backgrounds. The signal is predominantly visible in the HT350 region,
especially for LM1, which has larger squark and gluino masses.
There is, nevertheless, a sizeable signal in the HT300 control region as well.
This contamination of the control region by a potential signal is
taken into account in the limit calculation (Section~\ref{sec:yield-limit})
and both benchmark points are ruled out with a confidence level of 99\% or
higher (Section~\ref{sec:cmssm-limit}).

A variant of this background estimation method relies on the \RaT
measurement in the HT300 bin only and uses a small correction
from MC simulation to predict \RaT in the signal region. This variant
results in an estimate of $12.0\pm^{8.1}_{6.3}$ (stat) $\pm$0.4 (syst) events.
Another variant, also based on the independence of \RaT on \scalht when the data sample is
dominated by EWK processes, i.e.\ for $\alt>0.55$, uses the weighted
average of the \RaT values measured in the two control regions.
This value is then also used
in the signal region to obtain a background estimate of $12.5 \pm$1.9
(stat) $\pm$0.7 (syst). Within uncertainties, the three estimates
are in agreement. Furthermore, in the simulation all methods are
shown to provide an unbiased estimator of the number of total
background events.
For the remainder of this Letter the result of the first method,
which relies entirely on measured data and makes the most conservative
assumption on the evolution of the double ratio with
\scalht, is used to estimate the total background.

\subsection{\texorpdfstring{\wj and \ttNew background}{W+jets and t t-bar background} \label{sec:w-bkg}}

A second background estimation method uses an independent selection of
\wmunu\ + jets events in the data in order to assess the contribution from
SM processes with genuine \met. The \wmunu\ + jets are selected
as described in~\cite{toppaper}, with an energetic and isolated
muon in the final state, and by requiring the transverse mass of the W
to be larger than 30\,GeV (to ensure a very pure sample originating from W + jets
and \ttNew).
The muons are required to be separated
from the jets in the event by a distance larger than 0.5 in the $(\eta,\phi)$
plane. Since $\alt > 0.55$ implies $\mht/\scalht > 0.4$,
only events with $\mht>140\gev$ are considered in the signal region (HT350).
In the lower \scalht regions, this requirement is scaled accordingly to $\mht>120$
($100$)\gev for HT300 (HT250).

In the HT350 region this selection yields $25$ events, in agreement
with the $ 29.4 \pm 1.4$ events predicted by the simulation. In the HT250
(HT300) region, $134$ ($52$) W candidates are reconstructed, in
agreement with the prediction of $135.5 \pm 3.2$ ($56.7 \pm 2.2$)
events. The fraction of \wmunu\ + jets events with $\alt>0.55$ in the data is
also in good agreement with the simulation: seven data events are found
in the signal region, compared with $5.9 \pm 0.6$ events predicted, whereas $32$
($12$) events in the data pass the $\alt>0.55$ requirement in the
HT250 (HT300) region, compared to $29.2 \pm 1.4$ ($11.1\pm 1.1$)
events expected.

The number of W + jets and \ttNew events satisfying the hadronic final state
selection of Section~\ref{sec:evtsel}, $N^{\rm W;\ had}_{\rm data}$,
can be estimated from the number of events in the muon sample,
$N^{{\rm W};\ \mu}_{\rm data}$, and the expected relative ratio of
these two types of events. The value of this ratio is taken from
Monte Carlo simulation, which yields $N^{\rm W;\ had}_{\rm data} =
N^{\rm W;\ had}_{\rm MC}/{N_{\rm MC}^{{\rm W};\ \mu}} \times
N^{{\rm W};\ \mu}_{\rm data} \approx 0.86 \times N_{\rm data}^{{\rm W};\ \mu}$.
The total background from
W + jets and \ttNew processes is thus estimated to be $6.1^{+2.8}_{-1.9}$
(stat) $\pm$ 1.8 (syst). Given the reliance on simulation
for the factor $N^{\rm W;\ had}_{\rm MC}/{N_{\rm MC}^{{\rm W};\
\mu}}$, conservative uncertainties on all the parameters entering
this ratio have been assigned. The systematic uncertainty is
estimated to be $30\%$ and is dominated by the uncertainty on the
efficiency for vetoing leptons.

\subsection{\texorpdfstring{\znunu + jets background}{Z to nu nu-bar+jets background} \label{sec:z-bkg}}

The remaining irreducible background stems from \znunu\ + jets
events. An estimate of this background can be obtained from
$\gamma$ + jets events, which have a larger production cross section but
kinematic properties similar to those of
\znunu\ + jets events when the photon is ignored~\cite{PAS-SUS-08-002}.
These $\gamma$ + jets events provide a measurement of the acceptance of
the $\alt>0.55$ requirement directly from data. The $\gamma$ + jets sample is selected
by requiring photons, i.e.\ localized electromagnetic depositions
satisfying very tight isolation criteria, with $\pt>100\gev$, $|\eta|<$ 1.45,
and with a distance in the $(\eta,\phi)$ plane to any jet larger than 1.0.
Subsequently, the photon is ignored and the same hadronic final state selection as
described in Section~\ref{sec:evtsel} is applied. As in
Section~\ref{sec:w-bkg}, \mht is required to exceed 140\gev. This
selection yields seven events in the data compared with $6.5 \pm 0.4$ expected
from simulation. The relative acceptances, together with the
appropriate ratio of cross sections for $\gamma$ + jets and
\znunu\ + jets, taken from simulation, are then used
to estimate the number of \znunu\ + jets events in the signal region, found
to be $N(\znunu\ +{\rm  jets})= 4.4^{+2.3}_{-1.6}(\rm{stat}) \pm 1.8 (\rm{syst})$.
The main systematic uncertainties arise from the ratio of
cross sections between $\gamma$ + jets and \znunu\ + jets in the simulation (30\%), the
efficiency for photon identification (20\%), and the purity of the
photon selection (20\%), which add up to $\approx 40\%$.

To check the validity of this uncertainty estimate, the number of
$\gamma$ + jets events can also be used to predict the number of ${\rm
W}$ + jets events. Ten ${\rm W}$ + jets events are observed with $250 < \scalht < 350$\,GeV, $N_{\rm jets} = 2$ and $\alpha_{\rm T}>0.55$, in
agreement with the prediction of $8.5 \pm 1.5 {\rm (stat)} \pm 2.6 {\rm
(syst)}$ from the $\gamma$ + jets process. This agreement gives
confidence that the magnitude of the assigned systematic uncertainties is adequate.

As a further cross-check, the \wmunu\ + jets sample discussed above is
used to estimate the background from \znunu\ + jets events.
With the observed number of reconstructed \wmunu\ + jets events, the
ratio of cross sections and branching ratios for W and Z bosons, and the
ratio of reconstruction efficiencies estimated from simulation, $4.9^{+2.6}_{-1.8}$~(stat)
$\pm 1.5$~(syst) \znunu\ + jets events are predicted, in agreement with the value
obtained from the $\gamma$ + jets sample.

\subsection{Background estimate summary}

The SM backgrounds to this analysis have been evaluated with
independent data control samples.  First, from the lower \scalht
regions in the data, a prediction for the total SM background of
$9.4^{+4.8}_{-4.0}$ (stat) $\pm$1.0 (syst) events in the signal
region is obtained. Then, with a \wmunu\  + jet control
sample, a contribution of $6.1^{+2.8}_{-1.9}$ (stat) $\pm$ 1.8 (syst)
events from the combination of W + jets and \ttNew processes is
estimated. Finally, with the $\gamma$ + jets sample, the background
from \znunu\ + jets events is estimated to be $4.4^{+2.3}_{-1.6}$ (stat)
$\pm$ 1.8 (syst). Therefore, the estimate of the SM background arising
from EWK processes with genuine \met is $10.5^{+3.6}_{-2.5}$
events, which is in agreement with the inclusive estimate
obtained from the lower \scalht control regions. All these
background estimates are used in the limit calculation.

The potential effect of multiple interactions per bunch crossing (pileup)
in the event selection and in the background estimates is evaluated
by comparing the fraction \RaT, for several thresholds on \alt,
between events with only one primary vertex and events with more than
one primary vertex.  No discernible difference has been found.

\begin{figure}[ht]
 \begin{center}
    \includegraphics[width=0.45\linewidth]{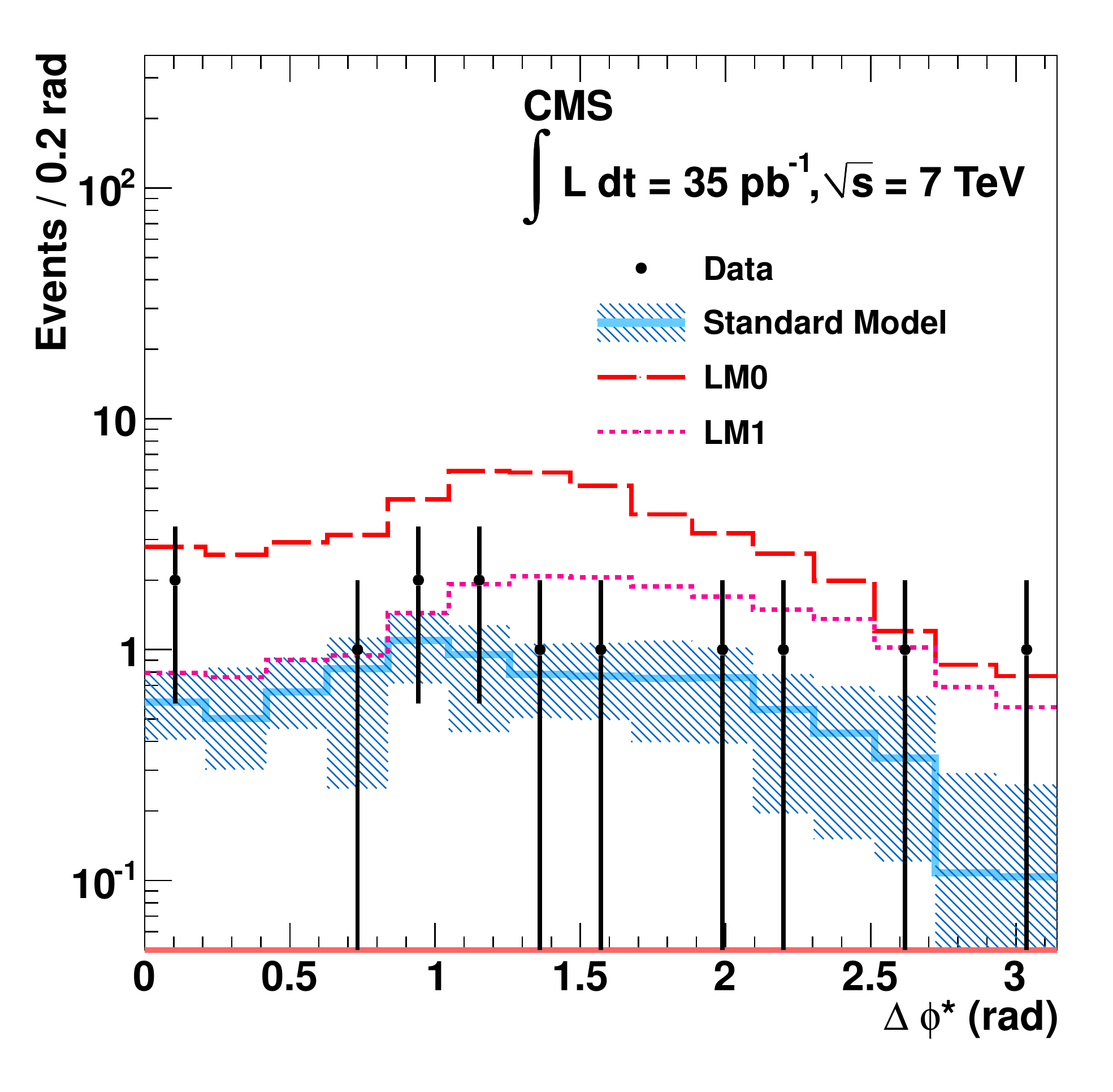}
    \includegraphics[width=0.45\linewidth]{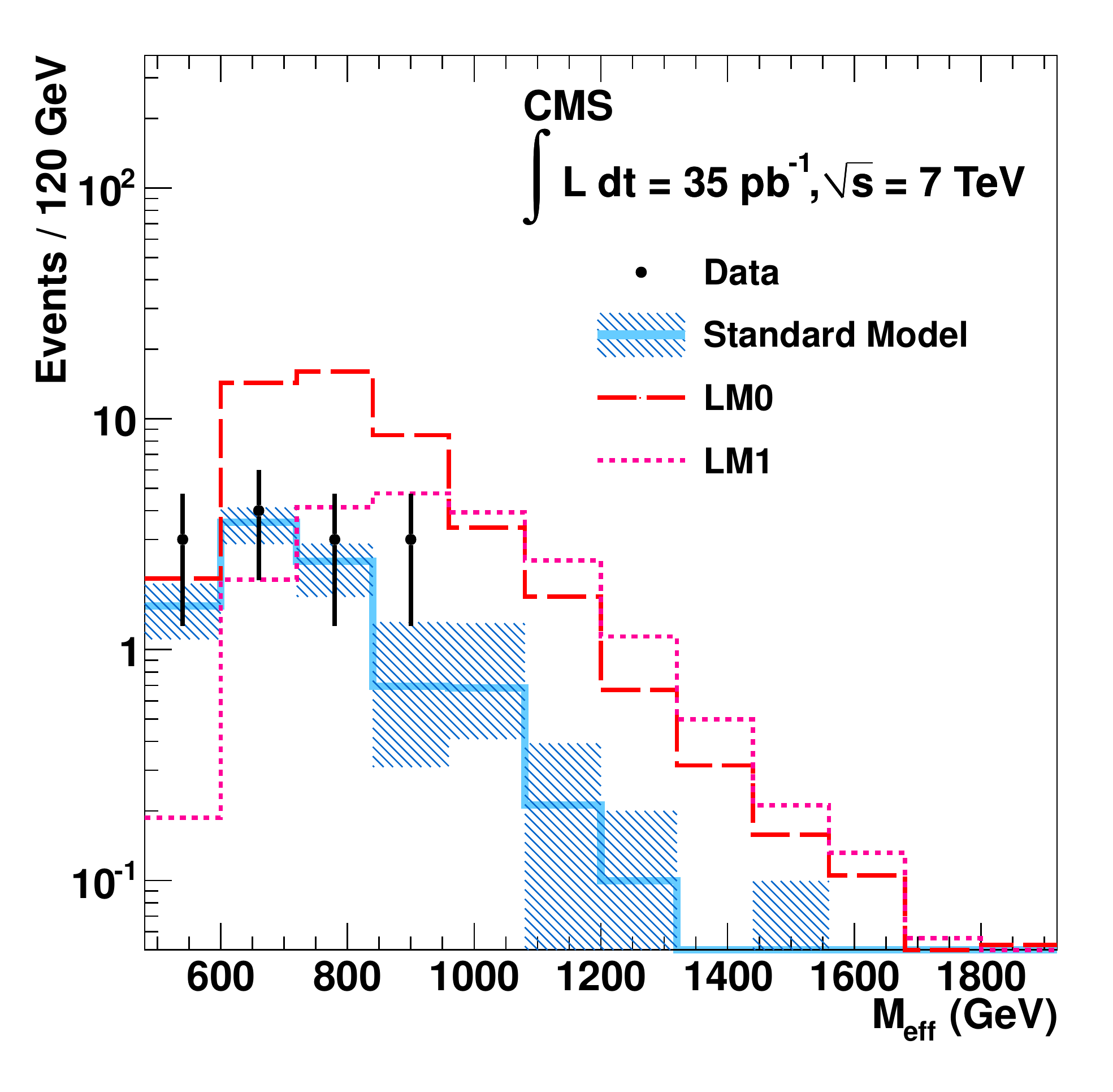}
\caption{\label{fig:dphistar} Left: the $\dphi^*$ distribution after all selection requirements. Right: the effective mass $M_{\rm eff}$ distribution after all selection requirements for SM processes and two low mass SUSY benchmark points. In both figures the hatched area corresponds to the uncertainty in the SM estimate as defined in Section~\ref{sec:hadsel}.}
 \end{center}
\end{figure}

The remaining events are found to exhibit the topological
and kinematic properties expected from the SM backgrounds.  Two distributions,
which are expected to show good separation between the SM background and
the SUSY signal, are shown in Fig.~\ref{fig:dphistar}.  The
$\dphi^*$ variable (Section~\ref{sec:finalsel}) is useful in identifying
mismeasured jets, since jet mismeasurements in QCD multijet events
result in small values of $\dphi^*$, whereas events with genuine \met,
e.g. from EWK processes, populate $\dphi^*$ evenly. In
Fig.~\ref{fig:dphistar} (left) the $\dphi^*$ distribution for the 13
data events which pass all selection requirements
is displayed. The data are consistent with EWK processes and there
is no indication of an enhanced contribution from QCD multijet processes
which would manifest itself at small values of $\dphi^*$.

The "effective mass'', $M_{\rm eff} = \scalht +
\mht$, which characterizes the overall energy scale of the event, is
shown in Fig.~\ref{fig:dphistar} (right) after all selection
requirements. The data are compared with the SM background
expectation along with two SUSY benchmark points. The shape and
magnitude of the $M_{\rm eff}$ distribution observed in the data are
consistent with the expectation from the SM backgrounds. The yields
expected from the LM0 and LM1 benchmark SUSY models are in excess of
the data over most of the $M_{\rm eff}$ range.

Both these variables exhibit differences
between SUSY signal events and events from SM backgrounds and could,
therefore, be used to improve the limits extracted in the following section.
We have chosen not to do so because the current search has been
optimized for the demonstration of a potential new signal, rather than for
the extraction of the most stringent limits in the SUSY parameter space.

In summary, 13 events are observed in the data, a yield consistent,
within the uncertainties, with the expectation from the
SM processes. In addition, the kinematic properties of
these events are consistent with the EWK backgrounds, with a negligible
contribution from QCD multijet processes.

\section{Interpretation of the Result}

\subsection{Method and limit on signal yield\label{sec:yield-limit}}

The background estimation methods described in the
previous section are combined to provide an estimate of the
total number of background events. This estimate is found to be compatible
with the number of events selected. An upper limit on the number of non-SM events consistent
with the measurements is derived using the Feldman-Cousins method~\cite{refFC},  which is generalized to 
take into account nuisance parameters by using the Profile Likelihood ratio~\cite{ProfileLikelihood}.
The input to the profiling method is the total likelihood function
$L_{\rm total}$ for the measurements in the control and signal regions.  To
construct this likelihood function, the numbers of events observed in the
signal region and in each of the control samples are treated as
independent event-counting experiments. The total likelihood function
can be written as

\begin{equation}
L_{\rm total} = L_{\rm signal} \cdot L_{\rm inclusive} \cdot L_{\rm W/\ttNew}\cdot L_{\znunu},
\end{equation}
where the different factors correspond to the likelihood of the measurement in 
the signal region, $L_{\rm signal}$, of the inclusive background extraction method using 
all three \scalht bins defined in Section~\ref{sec:inclusive-bkg}, $L_{\rm inclusive}$, of the exclusive measurement 
of  \ttNew and W + jet  event background described in Section~\ref{sec:w-bkg}, $L_{\rm W/\ttNew}$, and 
of the  \znunu\  background component described in Section~\ref{sec:z-bkg}, $L_{\znunu}$.

The likelihood functions are taken as Poisson-distributed probabilities  
to measure the number of events observed while expecting to see the 
estimated number of background events plus a certain fraction of CMSSM 
SUSY signal events. 
The expected backgrounds in the auxiliary measurements and in the signal-like 
region are related as described in Section~\ref{sec:background}. 
The ratio of the expected signal events in the control region and in the 
signal region is model-dependent and varies from point to point in the 
CMSSM parameter space.

The expected number of background events and the systematic uncertainties 
on the background prediction and on the signal selection efficiency
are treated as nuisance parameters. The probability density functions 
which describe the systematic uncertainties are assumed to be Gaussian with 
variance given by the systematic uncertainties derived in the previous section. 

The systematic uncertainties on the signal event yield can be split
into two parts: theoretical uncertainties on the predicted
cross section of the different production processes (squark-squark,
squark-gluino, gluino-gluino) and experimental uncertainties on
the integrated luminosity~\cite{CMSluminosity} and on the selection efficiency.

Whereas the theoretical uncertainties are strongly model dependent, the
experimental uncertainties are found to be essentially independent of the signal model. 
The experimental systematic uncertainties on the estimated event yield are the uncertainty on the
luminosity measurement (11\%), the effect of rejecting
events with jets pointing to masked ECAL regions (3\%), the modelling of
the lepton and photon vetoes in the simulation (2.5\%), and the effect of 
the uncertainty in the jet energy scale and resolution on the selection 
efficiency (2.5\%). These
uncertainties are included in the limit calculation.
The effect of multiple interactions per bunch crossing on the signal is evaluated 
by comparing the efficiency for signal events passing all selection requirements 
with and without the inclusion of multiple interactions in the simulation.
The effect on the efficiency is negligible.

If a potential signal contamination in the background control
samples is ignored, an upper limit on the number of signal events compatible with
the observations at 95\% confidence level (CL) can be obtained. For an
integrated luminosity of 35 pb$^{-1}$ this number is 13.4 events. 
The $p$ value for the hypothesis of standard model background only, 
calculated from the ratio of likelihoods, is 0.3.

\subsection{Interpretation within the CMSSM\label{sec:cmssm-limit}}

To interpret the consistency of the observed number of events with the
background expectation in the context of a model, and also to
facilitate the comparison with previous experimental results, an
exclusion limit in the CMSSM is set. This limit is obtained by testing, for
each point in the parameter space, whether the number of signal events
predicted after all selection requirements is compatible with observations 
at 95\% CL. 

Signal contamination in the data control samples used to estimate the 
background is also taken into account by explicitly including the number of 
signal expected in the control regions.
As the search is designed for robustness and background control, the same 
selection is applied at each point in the parameter space, and no dedicated
optimization for the CMSSM parameter space is performed.

An example of the analysis efficiency and corresponding event yields
after all selection requirements, broken down by the most relevant
production processes (squark-squark ($\sQuanew\sQuanew$), squark-gluino
($\sQuanew\sGlunew$), and gluino-gluino ($\sGlunew\sGlunew$)), is presented in
Table~\ref{tab:eventyields} for the benchmark point LM1. Two different
experimental efficiencies $\epsilon_{\rm total}$ and $\epsilon_{\rm signature}$ are given.
The first number, $\epsilon_{\rm total}$, is normalized to the total
number of signal events in LM1, while $\epsilon_{\rm signature}$ is
defined with respect to the total number of all-hadronic events in LM1
where, as in the analysis, leptons and photons are vetoed. For the
different production mechanisms, $\epsilon_{\rm total}$ varies from
12\% to 16\%. The signature-based efficiency is almost constant,
varying between 22\% and 23\%, which indicates that the analysis has a
uniform sensitivity to the different production channels in LM1.
With the current data, the LM1 and LM0 benchmark points are 
excluded at 99.2\% CL and 99.99\% CL, respectively.

\begin{figure}[h]
   \begin{center}
    \includegraphics[width=0.75\textwidth]{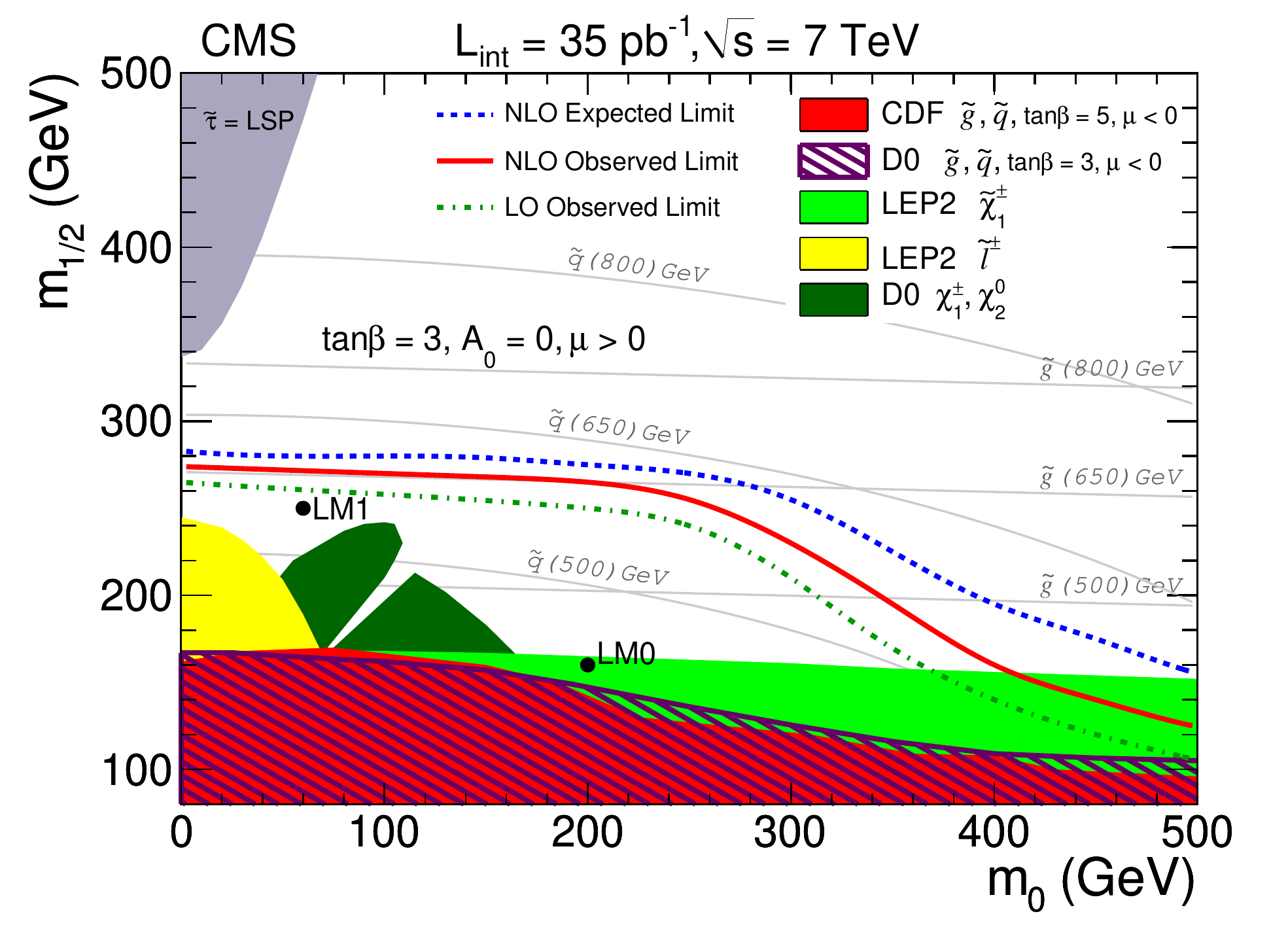}
 \caption{\label{fig:msugraexcl} Measured (red line) and expected (dashed blue line) $95\%$ CL exclusion contour at NLO in the CMSSM $(m_0, m_{1/2})$ plane for $\tan \beta = 3$, $A_0 = 0$ and $\sign(\mu) > 0$. The measured LO exclusion contour is shown as well (dot-dashed green line). The area below the curves is excluded by this measurement. Exclusion limits obtained from previous experiments are presented as filled areas in the plot. Grey lines correspond to constant squark and gluino masses. The plot also shows the two benchmark points LM0 and LM1 for comparison.}
\end{center}
\end{figure}

The 95\% CL limit in the $(m_0,m_{1/2})$ plane, for $\tan\beta=3$, $A_0=0$ and  $\sign(\mu) > 0$,
is shown in Fig.~\ref{fig:msugraexcl}. The SUSY particle spectrum is calculated
using SoftSUSY \cite{Allanach:2001kg}, and the signal events are
generated at leading order (LO) with \PYTHIA6.4.  Next-to-leading order
(NLO) cross sections, obtained with the program Prospino~\cite{Beenakker:1996ch}, 
are used to calculate the observed and expected exclusion contours. 
Systematic uncertainties on the NLO predictions due to the choice of the 
renormalization and factorization scales have been taken into account. The 
uncertainties on the used parton distribution functions (PDF) for CTEQ6.6~\cite{CTEQ66} 
are estimated from the envelope provided by the CTEQ6.6 error function. 
For reference, the observed limit using LO cross sections is also shown.

\begin{table}
  \caption{Breakdown of expected event yields and selection
    efficiencies for the most important production channels of the LM1 benchmark point after all selection requirements.
    No distinction has been made between \sQuanew and $\overline{\sQuanew}$. The quoted errors represent the statistical uncertainties on the yields and efficiencies. The efficiencies $\epsilon_{\rm total} $ and $\epsilon_{\rm signature}$ are defined in the text in Section~\ref{sec:cmssm-limit}.  }
\label{tab:eventyields}
\begin{center}
\begin{tabular}{|c|c|c|c|}
\hline
Production mechanism & Yields for 35 pb$^{-1}$ & $\epsilon_{\rm total}$(\%)  & $\epsilon_{\rm signature}$(\%) \\
\hline
\sQuanew\sQuanew & 9.7$\pm$0.1 & 16.0$\pm$0.1 & 22.2$\pm$0.4 \\
\sQuanew\sGlunew & 8.8$\pm$0.1 & 14.4$\pm$0.1 & 23.0$\pm$0.5 \\
\sGlunew\sGlunew & 0.71$\pm$0.02 & 12.0$\pm$0.4 & 22.5$\pm$2.0\\
\hline
\end{tabular}
\end{center}
\end{table}

The expected limit covers a larger part of the $(m_0$, $m_{1/2})$ plane than the
measured limit, as the number of events observed in the signal region is 
slightly larger than the number of background events predicted from the control regions. 
The excluded regions for the CDF search for jets + missing energy final 
states~\cite{CDFLimits} were obtained for $\tan\beta=5$, while those 
from D0~\cite{D0Limits} were obtained for $\tan\beta=3$, each with approximately 2~fb$^{-1}$ of data.
The LEP-excluded regions are based on searches for sleptons and
charginos~\cite{LEPLimits}. A comparison of the exclusion limit for
$\tan\beta=3$ to that for $\tan\beta=10$ for fixed values of $A_0=0$ and
$\sign(\mu) > 0$ indicates that the exclusion reach is only weakly
dependent on the value of $\tan\beta$; the limit shifts by less than $20\gev$ in
$m_0$ and by less than $10\gev$ in $m_{1/2}$.  
The D0 exclusion limit, valid for $\tan \beta = 3$  and  obtained from a search for associated production of charginos $\chi_1^{\pm}$ and neutralinos $\chi_2^{0}$ in trilepton final states~\cite{Abazov200934}, is also included in Fig.~\ref{fig:msugraexcl}. In contrast to the other limits presented in Fig.~\ref{fig:msugraexcl}, the result of the trilepton search is strongly dependent on the choice of  $\tan \beta$ and it reaches its highest sensitivity in the CMSSM for  $\tan \beta$ values below ten. 

\section{Summary}
The first search for supersymmetry in events collected by the CMS
experiment from proton-proton collisions at a centre-of-mass energy
of 7 TeV has been presented.  The final states with two or more
hadronic jets and significant missing transverse energy, as expected from
high-mass squark and gluino production and decays, have been analysed
in data corresponding to an integrated luminosity of 35~pb$^{-1}$. 
A search for a SUSY signal has been performed at high values of the scalar sum 
of the transverse energy of jets, \scalht.  The primary background, from QCD multijet events, has been reduced by
several orders of magnitude down to a negligible level using a robust
set of requirements designed specifically for the exploratory, early
data-taking phase of the experiment.  The sum of standard model
backgrounds has been estimated from an extrapolation of the data
observed at lower \scalht values.  The only remaining backgrounds have
been found to stem from electroweak processes, namely W + jet, Z + jet, and
\ttNew production, where the weak decays of the vector bosons involve
high-momentum neutrinos. An independent estimate of the electroweak
backgrounds, from \wmunu~+ jets decays as well as $\gamma$ + jets events
in the data together with input from simulation, has been found to be well
compatible with the estimate from control samples in data. Here, conservatively
large systematic uncertainties have been assigned to the background estimates. 
The measurements are in agreement with the expected
contributions from standard model processes. Limits on the CMSSM
parameters have been derived, and have been shown to
improve significantly those set by previous experiments.

\section{Acknowledgements}

We thank L. Dixon and the members of the Blackhat collaboration for discussions concerning 
vector-boson + jets production at the LHC. 
We wish to congratulate our colleagues in the CERN accelerator
departments for the excellent performance of the LHC machine. We thank
the technical and administrative staff at CERN and other CMS
institutes, and acknowledge support from: FMSR (Austria); FNRS and FWO
(Belgium); CNPq, CAPES, FAPERJ, and FAPESP (Brazil); MES (Bulgaria);
CERN; CAS, MoST, and NSFC (China); COLCIENCIAS (Colombia); MSES
(Croatia); RPF (Cyprus); Academy of Sciences and NICPB (Estonia);
Academy of Finland, ME, and HIP (Finland); CEA and CNRS/IN2P3
(France); BMBF, DFG, and HGF (Germany); GSRT (Greece); OTKA and NKTH
(Hungary); DAE and DST (India); IPM (Iran); SFI (Ireland); INFN
(Italy); NRF and WCU (Korea); LAS (Lithuania); CINVESTAV, CONACYT,
SEP, and UASLP-FAI (Mexico); PAEC (Pakistan); SCSR (Poland); FCT
(Portugal); JINR (Armenia, Belarus, Georgia, Ukraine, Uzbekistan); MST
and MAE (Russia); MSTD (Serbia); MICINN and CPAN (Spain); Swiss
Funding Agencies (Switzerland); NSC (Taipei); TUBITAK and TAEK
(Turkey); STFC (United Kingdom); DOE and NSF (USA).

\bibliography{auto_generated}   

\providecommand{\href}[2]{#2}\begingroup\raggedright\begin{thebibliography}{10}%
\makeatletter
\providecommand{\hrefCMSnoop }[0]{\@secondoftwo}%
\makeatother

\bibitem{ref:SUSY-1}
\hrefCMSnoop {} {Y.~A. Golfand and E.~P. Likhtman, ``Extension of the Algebra
  of {Poincare} Group Generators and Violation of p Invariance'',} \textit{
  JETP Lett.} \textbf{ 13} (1971) 323--326.

\bibitem{ref:SUSY0}
\hrefCMSnoop {} {J.~Wess and B.~Zumino, ``Supergauge transformations in four
  dimensions'',} \textit{ Nucl. Phys.} \textbf{ B70} (1974) 39.
  \href{http://dx.doi.org/10.1016/0550-3213(74)90355-1}{\texttt{
  doi:10.1016/0550-3213(74)90355-1}}.

\bibitem{ref:SUSY1}
\hrefCMSnoop {} {H.~P. Nilles, ``Supersymmetry, Supergravity and Particle
  Physics'',} \textit{ Phys. Reports} \textbf{ 110} (1984) 1.
  \href{http://dx.doi.org/10.1016/0370-1573(84)90008-5}{\texttt{
  doi:10.1016/0370-1573(84)90008-5}}.

\bibitem{ref:SUSY2}
\hrefCMSnoop {} {H.~Haber and G.~Kane, ``The Search for Supersymmetry: Probing
  Physics Beyond the Standard Model'',} \textit{ Phys. Reports} \textbf{ 117}
  (1987) 75. \href{http://dx.doi.org/10.1016/0370-1573(85)90051-1}{\texttt{
  doi:10.1016/0370-1573(85)90051-1}}.

\bibitem{ref:SUSY3}
\hrefCMSnoop {} {R.~Barbieri, S.~Ferrara, and C.~A. Savoy, ``Gauge Models with
  Spontaneously Broken Local Supersymmetry'',} \textit{ Phys. Lett.} \textbf{
  B119} (1982) 343.
  \href{http://dx.doi.org/10.1016/0370-2693(82)90685-2}{\texttt{
  doi:10.1016/0370-2693(82)90685-2}}.

\bibitem{ref:SUSY4}
\hrefCMSnoop {} {S.~Dawson, E.~Eichten, and C.~Quigg, ``Search for
  Supersymmetric Particles in Hadron - Hadron Collisions'',} \textit{ Phys.
  Rev.} \textbf{ D31} (1985) 1581.
  \href{http://dx.doi.org/10.1103/PhysRevD.31.1581}{\texttt{
  doi:10.1103/PhysRevD.31.1581}}.

\bibitem{ref:hierarchy1}
\hrefCMSnoop {} {E.~Witten, ``Dynamical Breaking of Supersymmetry'',} \textit{
  Nucl. Phys.} \textbf{ B188} (1981) 513.
  \href{http://dx.doi.org/10.1016/0550-3213(81)90006-7}{\texttt{
  doi:10.1016/0550-3213(81)90006-7}}.

\bibitem{ref:hierarchy2}
\hrefCMSnoop {} {S.~Dimopoulos and H.~Georgi, ``Softly Broken Supersymmetry and
  {SU(5)}'',} \textit{ Nucl. Phys.} \textbf{ B193} (1981) 150.
  \href{http://dx.doi.org/10.1016/0550-3213(81)90522-8}{\texttt{
  doi:10.1016/0550-3213(81)90522-8}}.

\bibitem{Farrar:1978xj}
\hrefCMSnoop {} {G.~R. Farrar and P.~Fayet, ``Phenomenology of the Production,
  Decay, and Detection of New Hadronic States Associated with Supersymmetry'',}
  \textit{ Phys. Lett.} \textbf{ B76} (1978) 575.
  \href{http://dx.doi.org/10.1016/0370-2693(78)90858-4}{\texttt{
  doi:10.1016/0370-2693(78)90858-4}}.

\bibitem{CDFLimits}
\hrefCMSnoop {} {{ CDF} Collaboration, ``Inclusive Search for Squark and Gluino
  Production in $p{\bar p}$ Collisions at $\sqrt{s} = 1.96$ {TeV}'',} \textit{
  Phys. Rev. Lett.} \textbf{ 102} (2009) 121801.
  \href{http://dx.doi.org/10.1103/PhysRevLett.102.121801}{\texttt{
  doi:10.1103/PhysRevLett.102.121801}}.

\bibitem{CDFtrileptons}
\hrefCMSnoop {} {{ CDF} Collaboration, ``Search for Supersymmetry in $p\bar{p}$
  Collisions at $\sqrt{s}=1.96$ {TeV} Using the Trilepton Signature for
  Chargino-Neutralino Production'',} \textit{ Phys. Rev. Lett.} \textbf{ 101}
  (2008), no.~25, 251801.
  \href{http://dx.doi.org/10.1103/PhysRevLett.101.251801}{\texttt{
  doi:10.1103/PhysRevLett.101.251801}}.

\bibitem{D0Limits}
\hrefCMSnoop {} {{ D0} Collaboration, ``Search for squarks and gluinos in
  events with jets and missing transverse energy using 2.1~$\mathrm{fb}^{-1}$
  of p$\mathrm{\bar{p}}$ collision data at $\sqrt{s} = 1.96~\TeV$'',} \textit{
  Phys. Lett.} \textbf{ B660} (2008) 449.
  \href{http://dx.doi.org/10.1016/j.physletb.2008.01.042}{\texttt{
  doi:10.1016/j.physletb.2008.01.042}}.

\bibitem{Abazov200934}
\hrefCMSnoop {} {{ D0} Collaboration, ``Search for associated production of
  charginos and neutralinos in the trilepton final state using 2.3~fb$^{-1}$ of
  data'',} \textit{ Phys. Lett.} \textbf{ B680} (2009) 34.
  \href{http://dx.doi.org/10.1016/j.physletb.2009.08.011}{\texttt{
  doi:10.1016/j.physletb.2009.08.011}}.

\bibitem{UA1Limits}
\hrefCMSnoop {} {{ UA1} Collaboration, ``Events with Large Missing Transverse
  Energy at the {CERN} Collider. 3. {M}ass Limits on Supersymmetric
  Particles'',} \textit{ Phys. Lett.} \textbf{ B198} (1987) 261.
  \href{http://dx.doi.org/10.1016/0370-2693(87)91509-7}{\texttt{
  doi:10.1016/0370-2693(87)91509-7}}.

\bibitem{UA2Limits}
\hrefCMSnoop {} {{ UA2} Collaboration, ``Search for Exotic Processes at the
  {CERN} p$\bar{\rm p}$ Collider'',} \textit{ Phys. Lett.} \textbf{ B195}
  (1987) 613. \href{http://dx.doi.org/10.1016/0370-2693(87)91583-8}{\texttt{
  doi:10.1016/0370-2693(87)91583-8}}.

\bibitem{HERALimits1}
\hrefCMSnoop {} {{ ZEUS} Collaboration, ``Search for stop production in
  {R}-parity-violating supersymmetry at {HERA}'',} \textit{ Eur. Phys. J.}
  \textbf{ C50} (2007) 269.
  \href{http://dx.doi.org/10.1140/epjc/s10052-007-0240-8}{\texttt{
  doi:10.1140/epjc/s10052-007-0240-8}}.

\bibitem{HERALimits2}
\hrefCMSnoop {} {{ H1} Collaboration, ``A Search for selectrons and squarks at
  {HERA}'',} \textit{ Phys. Lett.} \textbf{ B380} (1996) 461.
  \href{http://dx.doi.org/10.1016/0370-2693(96)00640-5}{\texttt{
  doi:10.1016/0370-2693(96)00640-5}}.

\bibitem{LEPLimits}
{ ALEPH, DELPHI, L3 and OPAL} Collaboration, \href
  {http://lepsusy.web.cern.ch/lepsusy} {LEPSUSYWG, ``Joint {SUSY} Working
  Group'',}. {LEPSUSYWG/02-06-2}.

\bibitem{ref:CMSSM}
\hrefCMSnoop {} {G.~L. Kane, C.~F. Kolda, L.~Roszkowski{ et~al.}, ``{Study of
  constrained minimal supersymmetry}'',} \textit{ Phys. Rev.} \textbf{ D49}
  (1994) 6173. \href{http://dx.doi.org/10.1103/PhysRevD.49.6173}{\texttt{
  doi:10.1103/PhysRevD.49.6173}}.

\bibitem{ref:MSUGRA}
\hrefCMSnoop {} {A.~H. Chamseddine, R.~Arnowitt, and P.~Nath, ``Locally
  Supersymmetric Grand Unification'',} \textit{ Phys. Rev. Lett.} \textbf{ 49}
  (Oct, 1982) 970--974.
  \href{http://dx.doi.org/10.1103/PhysRevLett.49.970}{\texttt{
  doi:10.1103/PhysRevLett.49.970}}.

\bibitem{PAS-SUS-09-001}
\href {http://cdsweb.cern.ch/record/1194509} {{ CMS} Collaboration, ``Search
  strategy for exclusive multi-jet events from supersymmetry at {CMS}'',}
  \textit{ CMS Physics Analysis Summary} \textbf{ SUS-09-001} (2009).

\bibitem{ref:CMS}
\hrefCMSnoop {} {{ CMS} Collaboration, ``The {CMS} experiment at the {CERN
  LHC}'',} \textit{ JINST} \textbf{ 03} (2008) S08004.
  \href{http://dx.doi.org/10.1088/1748-0221/3/08/S08004}{\texttt{
  doi:10.1088/1748-0221/3/08/S08004}}.

\bibitem{PAS-TRK-10-005}
\href {http://cdsweb.cern.ch/record/1279383} {{ CMS} Collaboration, ``Tracking
  and Primary Vertex Results in First 7 {TeV} Collisions'',} \textit{ CMS
  Physics Analysis Summary} \textbf{ TRK-10-005} (2010).

\bibitem{ref:antikt}
\hrefCMSnoop {} {M.~Cacciari, G.~P. Salam, and G.~Soyez, ``The anti-kt jet
  clustering algorithm'',} \textit{ JHEP} \textbf{ 0804:063} (2008).
  \href{http://dx.doi.org/10.1088/1126-6708/2008/04/063}{\texttt{
  doi:10.1088/1126-6708/2008/04/063}}.

\bibitem{PAS-JME-10-010}
\href {http://cdsweb.cern.ch/record/1308178} {{ CMS} Collaboration, ``Jet
  Energy Corrections determination at 7 {TeV}'',} \textit{ CMS Physics Analysis
  Summary} \textbf{ JME-10-010} (2010).

\bibitem{JME-09-008}
\href {http://cdsweb.cern.ch/record/1259924} {{CMS Collaboration},
  ``Calorimeter Jet Quality Criteria for the First {CMS} Collision Data'',}
  \textit{ CMS Physics Analysis Summary} \textbf{ JME-09-008} (2009).

\bibitem{PAS-EGM-10-004}
\href {http://cdsweb.cern.ch/record/1299116} {{ CMS} Collaboration, ``Electron
  reconstruction and identification at $\sqrt{s}$ = 7 {TeV}'',} \textit{ CMS
  Physics Analysis Summary} \textbf{ EGM-10-004} (2010).

\bibitem{PAS-MUO-10-002}
\href {http://cdsweb.cern.ch/record/1279140} {{ CMS} Collaboration,
  ``Performance of muon identification in pp collisions at $\sqrt{s}$ = 7
  {TeV}'',} \textit{ CMS Physics Analysis Summary} \textbf{ MUO-10-002} (2010).

\bibitem{PAS-EGM-10-005}
\href {http://cdsweb.cern.ch/record/1279143} {{ CMS} Collaboration, ``Photon
  reconstruction and identification at $\sqrt{s}$ = 7 {TeV}'',} \textit{ CMS
  Physics Analysis Summary} \textbf{ EGM-10-005} (2010).

\bibitem{pythia}
\hrefCMSnoop {} {T.~Sj\"ostrand, S.~Mrenna, and P.~Z. Skands, ``{PYTHIA} 6.4
  Physics and Manual; v6.420, tune {D6T}'',} \textit{ JHEP} \textbf{ 05} (2006)
  026. \href{http://dx.doi.org/10.1088/1126-6708/2006/05/026}{\texttt{
  doi:10.1088/1126-6708/2006/05/026}}.

\bibitem{Z2}
\hrefCMSnoop {} {R.~Field, ``Early {LHC} Underlying Event Data - Findings and
  Surprises'',} (2010).
\href{http://www.arXiv.org/abs/1010.3558}{\texttt{ arXiv:1010.3558}}.

\bibitem{madgraph}
\hrefCMSnoop {} {J.~Alwall {et~al.}, ``{MadGraph/MadEvent} v4: The New Web
  Generation'',} \textit{ JHEP} \textbf{ 09} (2007) 028.
  \href{http://dx.doi.org/10.1088/1126-6708/2007/09/028}{\texttt{
  doi:10.1088/1126-6708/2007/09/028}}.

\bibitem{PAS-SUS-08-005}
\href {http://cdsweb.cern.ch/record/1149915} {{ CMS} Collaboration, ``{SUSY}
  searches with dijet events'',} \textit{ CMS Physics Analysis Summary}
  \textbf{ SUS-08-005} (2008).

\bibitem{Randall:2008rw}
\hrefCMSnoop {} {L.~Randall and D.~Tucker-Smith, ``Dijet Searches for
  Supersymmetry at the {LHC}'',} \textit{ Phys. Rev. Lett.} \textbf{ 101}
  (2008) 221803.
  \href{http://dx.doi.org/10.1103/PhysRevLett.101.221803}{\texttt{
  doi:10.1103/PhysRevLett.101.221803}}.

\bibitem{PAS-JME-10-005}
\href {http://cdsweb.cern.ch/record/1294501} {{ CMS} Collaboration, ``{CMS}
  {MET} Performance in Events Containing Electroweak Bosons from pp Collisions
  at $\sqrt{s}= 7$ {TeV}'',} \textit{ CMS Physics Analysis Summary} \textbf{
  JME-10-005} (2010).

\bibitem{pythia8}
\hrefCMSnoop {} {T.~Sj\"ostrand, S.~Mrenna, and P.~Z. Skands, ``A Brief
  Introduction to {PYTHIA 8.1}'',} \textit{ Comput. Phys. Commun.} \textbf{
  178} (2008) 852. \href{http://dx.doi.org/10.1016/j.cpc.2008.01.036}{\texttt{
  doi:10.1016/j.cpc.2008.01.036}}.

\bibitem{toppaper}
\hrefCMSnoop {} {{ CMS} Collaboration, ``First Measurement of the Cross Section
  for Top-Quark Pair Production in Proton-Proton Collisions at sqrt(s)=7
  {TeV}'',} \textit{ Phys. Lett.} \textbf{ B695} (2011) 424.
  \href{http://dx.doi.org/10.1016/j.physletb.2010.11.058}{\texttt{
  doi:10.1016/j.physletb.2010.11.058}}.

\bibitem{PAS-SUS-08-002}
\href {http://cdsweb.cern.ch/record/1194471} {{ CMS} Collaboration,
  ``Data-Driven Estimation of the Invisible {Z} Background to the {SUSY MET}
  Plus Jets Search'',} \textit{ CMS Physics Analysis Summary} \textbf{
  SUS-08-002} (2008).

\bibitem{refFC}
\hrefCMSnoop {} {G.~J. Feldman and R.~D. Cousins, ``A Unified Approach to the
  Classical Statistical Analysis of Small Signals'',} \textit{ Phys. Rev.}
  \textbf{ D57} (1998) 3873--3889,
  \href{http://www.arXiv.org/abs/physics/9711021}{\texttt{
  arXiv:physics/9711021}}.
\href{http://dx.doi.org/10.1103/PhysRevD.57.3873}{\texttt{
  doi:10.1103/PhysRevD.57.3873}}.

\bibitem{ProfileLikelihood}
T.~A. Severini, ``Likelihood methods in statistics''.
\newblock Oxford University Press, 2000.

\bibitem{CMSluminosity}
\href {http://cdsweb.cern.ch/record/1279145} {{ CMS} Collaboration,
  ``Measurement of {CMS} luminosity'',} \textit{ CMS Physics Analysis Summary}
  \textbf{ EWK-10-004} (2010).

\bibitem{Allanach:2001kg}
\hrefCMSnoop {} {B.~C. Allanach, ``{SOFTSUSY}: a program for calculating
  supersymmetric spectra'',} \textit{ Comput. Phys. Commun.} \textbf{ 143}
  (2002) 305. \href{http://dx.doi.org/10.1016/S0010-4655(01)00460-X}{\texttt{
  doi:10.1016/S0010-4655(01)00460-X}}.

\bibitem{Beenakker:1996ch}
\hrefCMSnoop {} {W.~Beenakker, R.~Hopker, M.~Spira{ et~al.}, ``Squark and
  gluino production at hadron colliders'',} \textit{ Nucl. Phys.} \textbf{
  B492} (1997) 51--103.
  \href{http://dx.doi.org/10.1016/S0550-3213(97)00084-9}{\texttt{
  doi:10.1016/S0550-3213(97)00084-9}}.

\bibitem{CTEQ66}
\hrefCMSnoop {} {P.~M. Nadolsky {et~al.}, ``Implications of {CTEQ} global
  analysis for collider observables'',} \textit{ Phys. Rev.} \textbf{ D78}
  (2008) 013004. \href{http://dx.doi.org/10.1103/PhysRevD.78.013004}{\texttt{
  doi:10.1103/PhysRevD.78.013004}}.

\end{thebibliography}\endgroup

\cleardoublepage\appendix\section{The CMS Collaboration \label{app:collab}}\begin{sloppypar}\hyphenpenalty=5000\widowpenalty=500\clubpenalty=5000\textbf{Yerevan Physics Institute,  Yerevan,  Armenia}\\*[0pt]
V.~Khachatryan, A.M.~Sirunyan, A.~Tumasyan
\vskip\cmsinstskip
\textbf{Institut f\"{u}r Hochenergiephysik der OeAW,  Wien,  Austria}\\*[0pt]
W.~Adam, T.~Bergauer, M.~Dragicevic, J.~Er\"{o}, C.~Fabjan, M.~Friedl, R.~Fr\"{u}hwirth, V.M.~Ghete, J.~Hammer\cmsAuthorMark{1}, S.~H\"{a}nsel, C.~Hartl, M.~Hoch, N.~H\"{o}rmann, J.~Hrubec, M.~Jeitler, G.~Kasieczka, W.~Kiesenhofer, M.~Krammer, D.~Liko, I.~Mikulec, M.~Pernicka, H.~Rohringer, R.~Sch\"{o}fbeck, J.~Strauss, A.~Taurok, F.~Teischinger, P.~Wagner, W.~Waltenberger, G.~Walzel, E.~Widl, C.-E.~Wulz
% ...
\vskip\cmsinstskip
% ...
1:~~Also at CERN, European Organization for Nuclear Research, Geneva, Switzerland\\
% ...

\end{sloppypar}
\end{document}